\documentclass[11pt]{article}
\usepackage{amsfonts}
 \usepackage{amssymb}
 \usepackage{amsmath}
 \newfont{\bbbold}{msbm10}

 \newfont{\goth}{eufm10 scaled \magstep1}

 \def\d{\delta}\def\D{\Delta}

 \def\m{\mu}

 \def\be{\begin{equation}}\def\ee{\end{equation}}
 \def\bea{\begin{eqnarray}}\def\eea{\end{eqnarray}}
 \def\ba{\begin{array}}\def\ea{\end{array}}

 \def\del{\partial}

 \def\str{\rm str}

 \def\del{\partial}

 \def\3dt{\dot{3}}

 {}

 \def\nn{\nonumber}
 \def\bd{\begin{document}}
 \def\ed{\end{document}}
 \def\bea{\begin{eqnarray}}
 \def\ba{\begin{array}}\def\ea{\end{array}}
 \def\eea{\end{eqnarray}}
 \def\ft#1#2{{\textstyle{{\scriptstyle #1}\over {\scriptstyle #2}}}}
 \def\fft#1#2{{#1 \over #2}}
 \newcommand{\eq}[1]{(\ref{#1})}
 \def\eqs#1#2{(\ref{#1}-\ref{#2})}
 \def\det{{\rm det\,}}
 \def\tr{{\rm tr}}\def\Tr{{\rm Tr}}
  \def\str{{\rm str}} \def\diag{{\rm diag}}
 \def\sdet{{\rm sdet}}\def\symtr{{\rm symtr}}

\newcommand{\hoch}[1]{$^{#1}$}
\usepackage{epsfig}

\newcommand{\al}{\alpha}
\newcommand{\Be}{\mbox{B}}
\newcommand{\gm}{\gamma}
\newcommand{\Gm}{\Gamma}
\newcommand{\dl}{\delta}
\newcommand{\Dl}{\Delta}
\newcommand{\eps}{\epsilon}
\newcommand{\ep}{\epsilon}
\newcommand{\kp}{\kappa}
\newcommand{\ka}{\kappa}
\newcommand{\lm}{\lambda}
\newcommand{\Lm}{\Lambda}
\newcommand{\om}{\omega}
\newcommand{\pr}{\partial}
\newcommand{\pa}{\partial}
\newcommand{\dd}{\mbox{d}}
\newcommand{\dr}{{\rm d}}
\newcommand{\la}{\langle}
\newcommand{\ra}{\rangle}
\newcommand{\MS}{\mbox{MS}}

\begin{document}
\begin{titlepage}

\date{}
\vspace{-2.0cm}
\title{
{\vspace*{-6mm} 
\begin{flushright}
\small{LAPTH-1159/06 }
\end{flushright}
\vspace{10mm}
}
Magic identities for conformal four-point integrals
\vspace*{0mm}}
\author{J.M.\ Drummond$^{a}$,\quad J.\ Henn$^{a}$,\quad
  V.A. Smirnov$^{b}$,\quad E.\ Sokatchev$^{a}$ \\[5mm] 
   {\small $^a$   Laboratoire d'Annecy-le-Vieux de
Physique Th\'{e}orique  LAPTH\footnote{UMR 5108 associ\'{e}e \`{a}
 l'Universit\'{e} de Savoie}}\\
  {\small  B.P. 110,  F-74941 Annecy-le-Vieux,
  France }\\
 {\small $^b$  Nuclear Physics Institute of Moscow State University}\\
 {\small Moscow 119992, Russia} }
\maketitle
\vspace*{0mm}

\abstract
{We propose an iterative procedure for constructing
classes of off-shell four-point conformal integrals which are
identical. The proof of the identity is based on the conformal
properties of a subintegral common for the whole class. The
simplest example are the so-called `triple scalar box' and `tennis
court' integrals. In this case we also give an independent proof using
the method of Mellin--Barnes representation which can be applied
in a similar way for general off-shell Feynman integrals.}

\vspace{2truecm}
\thispagestyle{empty}
\end{titlepage}

\vfill
\newpage

\section{Introduction}
\label{sec:INTRO}

Four-point correlators in the ${\cal N}=4$ super-Yang-Mills
conformal field theory have attracted considerable attention since
the formulation of the AdS/CFT conjecture \cite{Maldacena}. They
can provide non-trivial dynamical information about the CFT side
of the correspondence, which can then be compared to its AdS dual.
In particular, the correlators of four `protected' stress-tensor
multiplets have been extensively studied. It has been found that
their form is more restricted than would follow from just
superconformal kinematics. This property, called `partial
non-renormalisation' in \cite{Eden:2000bk} is observed in the
perturbative one-loop \cite{Gonzalez-Rey:1998tk} and two-loop
\cite{Eden:2000mv} CFT calculations, as well as in their AdS
supergravity (or strong coupling) dual \cite{Arutyunov:2000py}.
These explicit results have been analysed through OPE methods
\cite{Arutyunov:2000ku} and the two-loop anomalous dimensions of
all twist two operators in the theory were found in
\cite{Dolan:2004iy}. In all these studies conformal four-point
integrals have been instrumental.

In a parallel development, on-shell four-gluon planar scattering
amplitudes in ${\cal N}=4$ SYM have been investigated in
\cite{Anastasiou:2003kj} and a remarkable conjecture about their
iterative structure has been made, based on the comparison of one-
and two-loop results. The conjecture was confirmed at three loops
in \cite{Bern:2005iz}. If true to all orders, this iterative
property may allow the resummation of the perturbative series and
may be the manifestation of some form of integrability of the
theory. One of the results of \cite{Bern:2005iz} was the large
spin asymptotic value of the anomalous dimension of twist two
operators, in agreement with the conjectured three-loop formula of
\cite{Kotikov:2004er}. The latter also received impressive
confirmation from the integrable model proposed in
\cite{Staudacher:2004tk}.

Although it may seem that the two problems, that of the
correlators of gauge-invariant composites and that of gluon
scattering amplitudes, are unrelated, it is quite significant that
in both studies one deals with the same conformal four-point
integrals. Up to two loops, these are the so-called `scalar box'
(or `ladder') integrals.\footnote{The off-shell ladder integrals
in four dimensions for an arbitrary number of loops have been
evaluated in \cite{Usyukina:1992jd,Usyukina:1993ch} and
generalised to arbitrary dimensions in \cite{Isaev:2003tk}.} At
three loops, in addition to the triple scalar box a new integral
named `tennis court' has appeared in \cite{Bern:2005iz}. In the
context of the scattering amplitudes these two integrals are put
on the massless shell, whereby they become infrared and collinear divergent.
Their pole structure in dimensional regularisation is quite
different, as shown in \cite{Bern:2005iz}. In the present paper we
prove that the two integrals, considered {\it off shell}, are
identical. We
first show this by a very simple argument, based on a `turning
symmetry' property of the two-loop scalar box subintegral common
for both three-loop integrals. It should be stressed that our
proof requires conformal invariance in strictly four dimensions,
therefore it does not apply to the dimensionally regularised
on-shell version of the integrals. To rule out the possibility of
contact terms spoiling the proof we give an alternative argument which
relates the two three-loop integrals to the same four-loop integral
under the action of a differential operator.
We then present a simple graphical rule
for constructing identical integrals which is easy to iterate to
any number of loops. In some sense our iteration procedure (or
`slingshot rule') resembles the so-called `rung rule' of
\cite{Anastasiou:2003kj,Bern:2005iz}.  Thus, at four loops we
produce five apparently different, but in fact identical integrals
obtained by iterating the already established three-loop identity
of the scalar box and the tennis court. We then give an
independent confirmation of the latter by explicitly computing the
two integrals using the Mellin--Barnes method.

\section{Conformal four-point integrals}

We will discuss an infinite class of conformal four-point integrals in
four dimensions\footnote{In this section we consider and prove
  identities for Euclidean integrals. The corresponding Minkowskian
  version of the identities can be obtained through Wick rotation of the
  integrals. In the Euclidean context we consider integrals with
  separated external points, $x_{ij}\neq 0$. This is the Euclidean
  analogue of the off-shell regime, $x_{ij}^2 \neq 0$, for a
  Minkowskian integral.},
each of which is essentially described by a function of two
variables. We begin with the simplest example, the one-loop ladder
integral,
\be
h^{(1)}(x_1,x_2,x_3,x_4) = \int \frac{d^4 x_5}{x_{15}^2 x_{25}^2 x_{35}^2
x_{45}^2} = \frac{1}{x_{13}^2 x_{24}^2} \Phi^{(1)}(s,t).
\label{phi1}
\ee
Here $x_{ij} = x_i - x_j$ and the conformal cross-ratios $s$ and $t$
are
\be\label{cro}
s = \frac{x_{12}^2 x_{34}^2}{x_{13}^2 x_{24}^2}, \hspace{20pt} t =
\frac{x_{14}^2 x_{23}^2}{x_{13}^2 x_{24}^2}.
\ee

\begin{figure}[htbp]
\begin{center}
\ \psfig{file=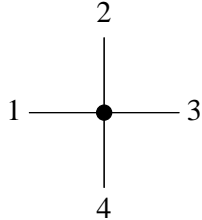}
\end{center}
\caption{\small The one-loop ladder integral. Each line represents a propagator
with the integration point given by a solid vertex. The reason for the
names ladder and box is clearer in the momentum representation of the
same integral.} 
\label{figure:1ladder}
\end{figure}

The fact that the integral is characterised by a single
function of two variables follows from its conformal covariance
\cite{Broadhurst:1993ib}. Indeed, performing a conformal inversion on
all points,
\be
x^\m \longrightarrow  \frac{x^\m}{x^2} \implies x_{ij}^2
\longrightarrow \frac{x_{ij}^2}{x_i^2 x_j^2},\hspace{10pt}
d^4 x_5 \longrightarrow \frac{d^4 x_5}{x_5^8} ,
\ee
we find that the integral transforms
covariantly with weight one at each point,
\be
h^{(1)}(x_1,x_2,x_3,x_4) \longrightarrow x_1^2 x_2^2 x_3^2 x_4^2
h^{(1)}(x_1,x_2,x_3,x_4) .
\ee
Since rotation and translation invariance are manifest, we
conclude
that the integral is given by a conformally covariant combination of
propagators multiplied by a function of the conformally
invariant cross-ratios (\ref{cro}).

The function $\Phi^{(1)}(s,t)$ has been calculated in
\cite{tHooft:1978xw,Usyukina:1992jd}, where it was also shown that the same
function appears in a three-point integral. The latter can be
obtained  from the four-point one by sending one of the points to
infinity \cite{Broadhurst:1993ib}. We can multiply equation
(\ref{phi1}) by $x_{13}^2$, say, and then take the limit $x_3
\longrightarrow \infty$. This gives,
\be
h_{\rm 3pt}^{(1)}(x_1,x_2,x_4) = \lim_{x_3 \rightarrow \infty} x_{13}^2
h^{(1)}(x_1,x_2,x_3,x_4) = \int \frac{d^4x_5}{x_{15}^2 x_{25}^2
  x_{45}^2} = \frac{1}{x_{24}^2} \Phi^{(1)}(\hat s,\hat t),
\ee
where the cross-ratios $s$ and $t$ have become $\hat s$ and $\hat t$ in the
limit,
\be
s\longrightarrow \hat s = \frac{x_{12}^2}{x_{24}^2}, \hspace{20pt}
t\longrightarrow \hat t = \frac{x_{14}^2}{x_{24}^2}.
\ee
Thus the three-point integral contains the same information as the
four-point integral, i.e. the same function of two variables. The
reason is that one can use translations and conformal inversion to
take the point $x_3$ to infinity and the function of the cross-ratios
is invariant under these transformations.


The integral (\ref{phi1}) is the first in an infinite series of conformal
integrals, the $n$-loop ladder (or scalar box) integrals, which have all
been evaluated \cite{Usyukina:1993ch}.
In particular the 2-loop ladder integral is given by
\be
h^{(2)}(x_1,x_2,x_3,x_4) = x_{24}^2 \int \frac{d^4 x_5 d^4 x_6}{x_{15}^2
  x_{25}^2 x_{45}^2 x_{56}^2 x_{26}^2 x_{46}^2 x_{36}^2} =
\frac{1}{x_{13}^2 x_{24}^2} \Phi^{(2)}(s,t).
\label{2ladder}
\ee
The prefactor $x_{24}^2$ is present to give conformal weight one at
each external point.
\begin{figure}[htbp]
\begin{center}
\ \psfig{file=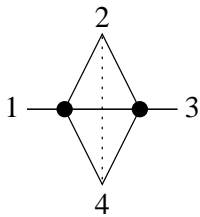}
\end{center}
\caption{\small The two-loop ladder integral. The dashed line represents the
  numerator $x_{24}^2$.}
\label{figure:2ladder}
\end{figure}

Again conformal transformations can be used to justify the appearance
of the 2-variable function $\Phi^{(2)}$.
The r.h.s. of (\ref{2ladder}) is invariant under the
pairwise swap $x_1 \longleftrightarrow x_2$,
$x_3 \longleftrightarrow x_4$, hence
\be
h^{(2)}(x_2,x_1,x_4,x_3)=h^{(2)}(x_1,x_2,x_3,x_4). \label{turning}
\ee
This symmetry is not immediately evident from the integral. It is its
conformal nature which allows this identification.
\begin{figure}[htbp]
\begin{center}
\ \psfig{file=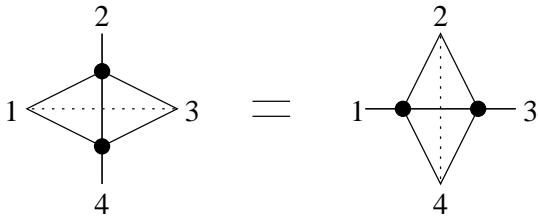}
\end{center}
\caption{\small The two-loop turning identity obtained from the
  pairwise point swap,
  $x_1 \longleftrightarrow x_2$, $x_3 \longleftrightarrow x_4$.}
\label{figure:turning}
\end{figure}

At three loops we consider two conformal integrals, the three-loop ladder,
\be
h^{(3)}(x_1,x_2,x_3,x_4) = x_{24}^4 \int \frac{d^4 x_5 d^4 x_6
  d^4 x_7}{x_{15}^2
  x_{25}^2 x_{45}^2 x_{56}^2 x_{26}^2 x_{46}^2 x_{67}^2 x_{27}^2 x_{47}^2
  x_{37}^2} = \frac{1}{x_{13}^2 x_{24}^2} \Phi^{(3)}(s,t),
\ee
and the so-called `tennis court' \cite{Bern:2005iz},
\be
g^{(3)}(x_1,x_2,x_3,x_4) = x_{24}^2 \int \frac{x_{35}^2\ d^4 x_5 d^4 x_6
  d^4 x_7
  }{x_{15}^2 x_{25}^2 x_{45}^2 x_{56}^2 x_{57}^2 x_{67}^2 x_{26}^2
  x_{47}^2 x_{36}^2 x_{37}^2} = \frac{1}{x_{13}^2 x_{24}^2}
\Psi^{(3)}(s,t) \label{tc}
\ee
\begin{figure}[htbp]
\begin{center}
\ \psfig{file=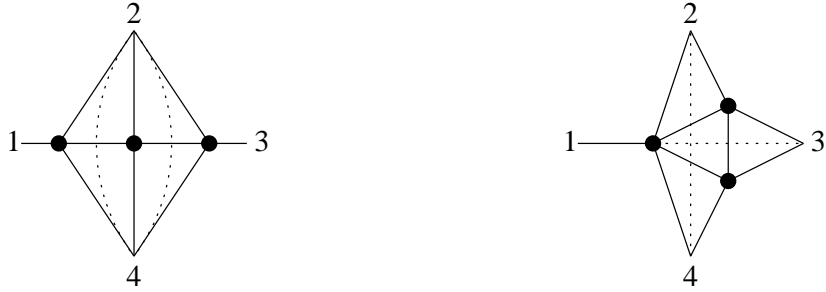}
\end{center}
\caption{\small Two examples of three-loop conformal four-point integrals, the
three-loop ladder and the `tennis-court'.}
\label{figure:3loop}
\end{figure}
Notice the presence of the numerator $x_{35}^2$ in the integrand
of the tennis court. It is needed to balance the conformal weight
of the five propagators coming out of point 5.

We will show that the three-loop ladder and the
tennis court are in fact the same, i.e. we will prove $\Phi^{(3)} =
\Psi^{(3)}$.  First we shall present a diagrammatic argument. We
consider the $n$-loop ladder as being iteratively constructed
from the $(n-1)$-loop ladder by integrating against a `slingshot' (the
`0-loop' ladder is a product of free propagators). For
example we write the three-loop ladder as
\be
h^{(3)}(x_1,x_2,x_3,x_4) = x_{24}^2 \int \frac{d^4x_5}{x_{15}^2 x_{25}^2
  x_{45}^2} \Bigl( x_{24}^2 \int \frac{d^4x_6 d^4x_7}{x_{56}^2 x_{26}^2
  x_{46}^2 x_{67}^2 x_{27}^2 x_{47}^2 x_{37}^2}\Bigr),
\ee
where inside the parentheses we recognise the two-loop ladder integral
(\ref{2ladder}).
\begin{figure}[htbp]
\begin{center}
\ \psfig{file=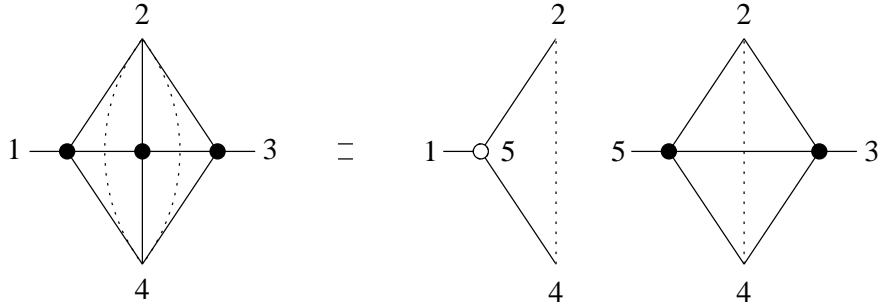}
\end{center}
\caption{\small The three-loop ladder expressed as the integral of the
  two-loop ladder against the `slingshot'. The empty vertex is the point $x_5$
  which must be identified with the point $x_5$ from the two-loop
  ladder sub-integral before being integrated over.}
\label{figure:slingshot}
\end{figure}

\begin{figure}[htbp]
\begin{center}
\ \psfig{file=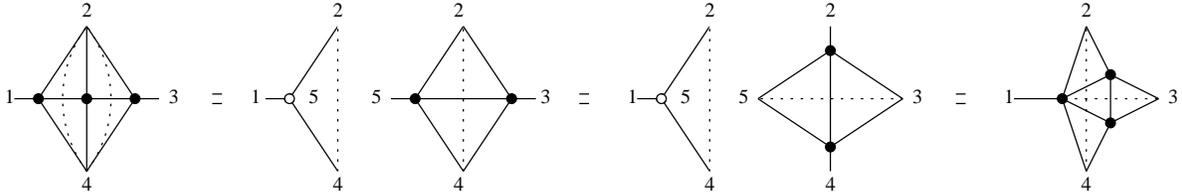}
\end{center}
\caption{\small Diagrammatic representation of the proof of equality of the
  tennis court and the three-loop ladder. The identity follows from the
  turning identity (\ref{turning}) for the two-loop subintegral.}
\label{figure:proof}
\end{figure}

We can then show the equality of the three-loop ladder and the tennis court by
using the turning symmetry (\ref{turning}) on the two-loop ladder
sub-integral. Then the tennis court integral (\ref{tc}) can be recognised as
the turned two-loop ladder integrated against the slingshot,
\begin{align}
h^{(3)}(x_1,x_2,x_3,x_4) &= x_{24}^2 \int \frac{d^4x_5}{x_{15}^2 x_{25}^2
  x_{45}^2} h^{(2)}(x_5,x_2,x_3,x_4), \notag \\
&= x_{24}^2 \int \frac{d^4x_5}{x_{15}^2 x_{25}^2 x_{45}^2}
  h^{(2)}(x_2,x_5,x_4,x_3), \notag \\
&= g^{(3)}(x_1,x_2,x_3,x_4). \label{proof}
\end{align}
This proof can be more easily seen in the diagram (Fig. \ref{figure:proof}).

In using the turning identity (\ref{turning}) we have ignored the
possibility of contact terms. These could, in principle, spoil the
derivation of identities like $\Phi^{(3)}=\Psi^{(3)}$ as the proof
(\ref{proof}) involves turning a subintegral. Contact terms could then
generate regular terms upon doing one further integration. We now give
an argument why this cannot happen for any conformal four-point
integral. We again use the example of the 3-loop ladder and tennis
court identity.

Consider inserting the $n$-loop subintegral (the 2-loop ladder in this
case) into an H-shaped frame with a dashed line across the top, as
illustrated below. This generates an $(n+2)$-loop integral which is
conformal with weight 1 at each external point (provided the
subintegral is conformal with weight 1 at each external point).

\begin{figure}[htbp]
\begin{center}
\ \psfig{file=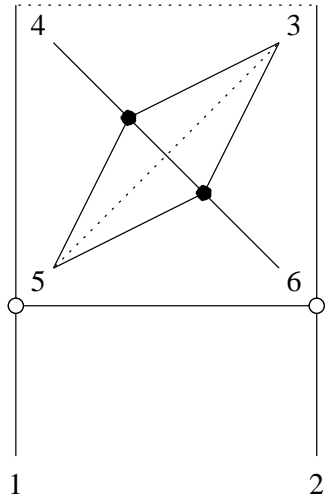}
\end{center}
\caption{\small The 2-loop ladder inserted into an H-shaped frame,
  generating a 4-loop integral.}
\label{figure:Hframe}
\end{figure}

When inserting the 2-loop ladder in this way the 4-loop integral one
obtains is
\begin{align}
f^{(4)}(x_1,x_2,x_3,x_4)&=
x_{34}^2 \int \frac{d^4x_5 d^4x_6} {x_{15}^2 x_{45}^2 x_{56}^2
  x_{26}^2 x_{36}^2} x_{35}^2 \int \frac{d^4x_7 d^4x_8}{x_{67}^2
  x_{57}^2 x_{37}^3 x_{78}^2 x_{58}^2 x_{38}^2 x_{48}^2} \notag \\
&= \frac{1}{x_{13}^2 x_{24}^2}f(s,t).
\label{f4int}
\end{align}
As usual, the second equality follows from conformality.

Now we consider the action of $\Box_1$ on the above integral using 

\be
\Box \frac{1}{x^2} = - 4 \pi^2 \d(x).
\ee

On the integral one obtains 

\be
-4\pi^2 \frac{x_{34}^2 x_{13}^2}{x_{14}^2} \int \frac{d^4x_6 d^4x_7
  d^4x_8}{x_{26}^2 
  x_{16}^2 x_{36}^2 x_{67}^2 x_{17}^2 x_{37}^2 x_{78}^2 x_{18}^2
  x_{38}^2 x_{48}^2} = -\frac{4 \pi^2 x_{34}^2}{x_{13}^4 x_{14}^2
  x_{24}^2} \Phi^{(3)}(s,t). 
\ee

On the functional form of (\ref{f4int}) one uses the chain rule to
derive the action of a differential operator on the function $f$. In
this way we find the differential equation,

\be
\frac{x_{23}^2 x_{34}^2}{x_{13}^6 x_{24}^4} \D^{(2)}_{st} f(s,t) =
-\frac{\pi^2 x_{34}^2}{x_{13}^4 x_{14}^2 x_{24}^2} \Phi^{(3)}(s,t).
\label{pdephi}
\ee

The operator $\D^{(2)}_{st}$ is given explicitly by

\be
\D^{(2)}_{st} = s \del_s^2 + t\del_t^2 + (s+t-1)\del_s \del_t +
2\del_s + 2\del_t.
\ee

Similarly we can act with $\Box_2$ on the 4-loop integral to obtain
the following integral,

\be
-4\pi^2 \frac{x_{34}^2}{x_{23}^2}\int \frac{d^4x_5 d^4x_7 d^4x_8
  x_{35}^2}{x_{15}^2 
  x_{25}^2 x_{45}^2 x_{57}^2 x_{58}^2 x_{78}^2 x_{27}^2 x_{48}^2
  x_{37}^2 x_{38}^2} = -\frac{4\pi^2 x_{34}^2}{x_{23}^2 x_{13}^2
  x_{24}^4} \Psi^{(3)}(s,t), 
\ee

and the corresponding differential equation,

\be
\frac{x_{14}^2 x_{34}^2}{x_{24}^6 x_{13}^4} \D^{(2)}_{st} f(s,t) =
-\frac{\pi^2 x_{34}^2}{x_{23}^2 x_{13}^2 x_{24}^4} \Psi^{(3)}(s,t).
\label{pdepsi}
\ee

From (\ref{pdephi},\ref{pdepsi}) it follows that
$\Phi^{(3)}=\Psi^{(3)}$, the point being that one obtains the {\it
  same} differential operator $\D^{(2)}_{st}$ under the two $\Box$
operations. The argument has the obvious generalisation of placing any
conformal integral (in any orientation) inside the frame. 
This argument indirectly shows that the previous argument
(\ref{proof}) based on turning the subintegral cannot suffer from
contact term contributions.

\begin{figure}[htbp]
\begin{center}
\ \psfig{file=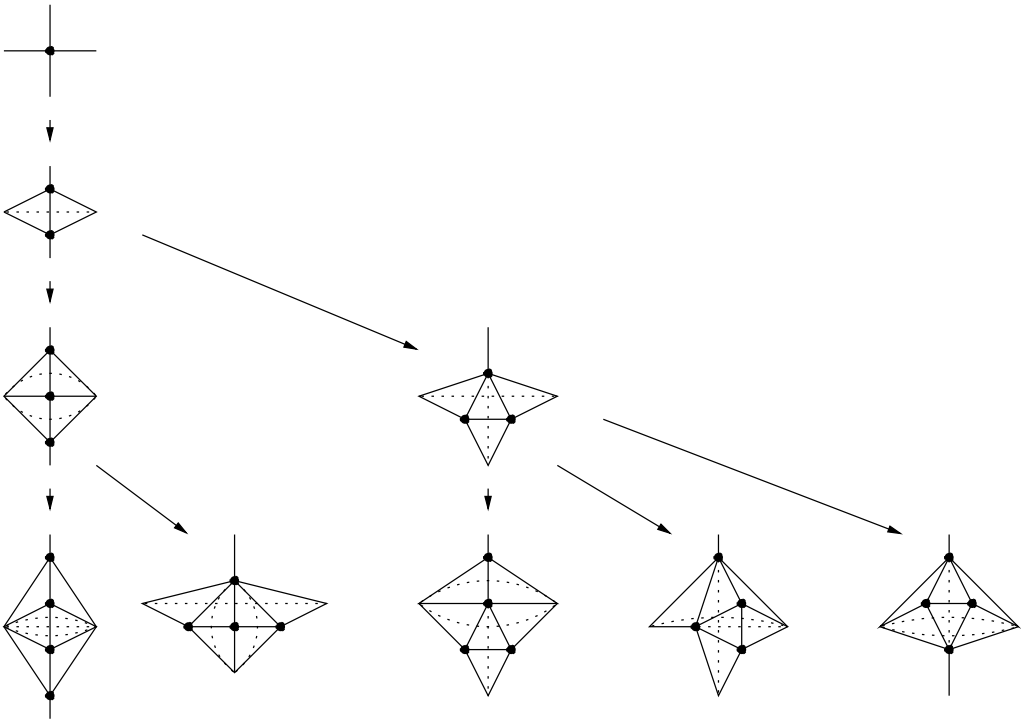}
\end{center}
\caption{\small The integrals in a given row are all equivalent. They
  generate the integrals in
the next row by being integrated in all possible orientations against
  the slingshot attached from above. The ladder series is in the
  left-most column.} 
\label{figure:cascadex}
\end{figure}

The identity we have obtained at three loops is just the first example
of an infinite set of identities which all come from the turning
symmetry of subintegrals. We generate $(n+1)$-loop
integrals by integrating $n$-loop integrals against the slingshot in
all possible orientations. The resulting integrals are equal by
turning identities of the form (\ref{turning}). At two loops we get
just one integral (the two-loop ladder). At three loops we have
already seen two equivalent integrals (ladder and tennis
court). At four loops we generate two equivalent integrals from the
three-loop ladder and three equivalent integrals from the tennis
court. Finally, all five four-loop integrals obtained in this way are
equivalent by the three-loop identity for the ladder and tennis court
(see Fig. \ref{figure:cascadex}).

\begin{figure}[htbp]
\begin{center}
\ \psfig{file=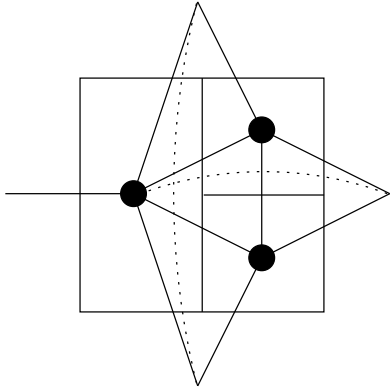}
\end{center}
\caption{\small The conversion from the momentum notation to the coordinate
  space notation. The pictures represent the same integral after a
  change of variables.}
\label{figure:notation}
\end{figure}

\begin{figure}[htbp]
\begin{center}
\ \psfig{file=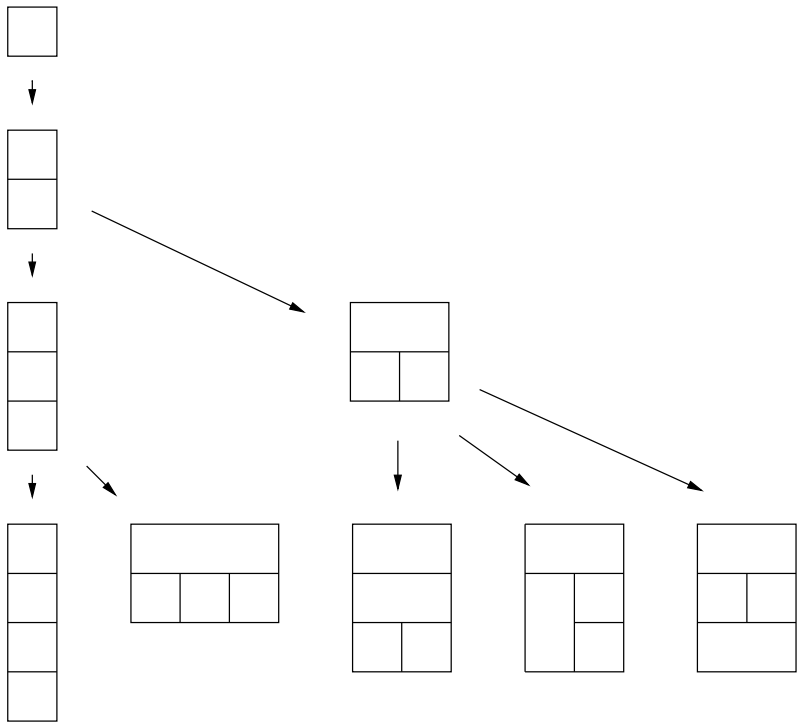}
\end{center}
\caption{\small The momentum notation for our integrals up to four
  loops. The slingshot translates into the top box in each
  diagram, beneath which are the integrals at one loop lower, arranged
  in all possible orientations. The ladder series is again in the
  left-most column.}
\label{figure:cascadep}
\end{figure}

In general it is more common to give the diagrams in the
`momentum' representation (which has nothing to do with the
Fourier transform) where we regard the integrations as integrals
over loop momenta rather than coordinate space vertices. This
representation is neater but the numerators need to be described
separately as they do not appear in the diagrams. To return to the
coordinate space integrals one places a vertex inside each loop
and connects them with propagators through each line. We show this
in Fig.  \ref{figure:notation} for the tennis court integral. The
momentum-space version of the four generations of integrals from
Fig. \ref{figure:cascadex} is then given in Fig.
\ref{figure:cascadep}.

\pagebreak

\section{Evaluating off-shell four-point Feynman integrals
by Mellin--Barnes representation}

Let us show how the above identity between the off shell triple box and tennis
court can straightforwardly be obtained by means of
the method of Mellin--Barnes (MB) representation.
This method is one of the most powerful
methods of evaluating individual Feynman integrals.\footnote{
It is especially successful for evaluating four-point Feynman
integrals.
For massless off-shell four-point integrals, first results were obtained by
means of MB representation in \cite{Usyukina:1992jd,Usyukina:1993ch}. In the
context of dimensional regularisation, with the space-time dimension
$d=4-2\ep$ as
a regularisation parameter, two alternative strategies
for resolving the structure of singularities in $\ep$ were
suggested in \cite{MB1,MB2} where first results on evaluating
four-point on-shell massless Feynman integrals were obtained.
Then these strategies
were successfully applied to
evaluate massless on-shell double \cite{MB1,MB2,GTGR0ATT,SV,AGORT}
and triple \cite{3b,Bern:2005iz} boxes, with results written in terms of
harmonic polylogarithms \cite{HPL}, double boxes with one leg off
shell \cite{S2} and massive on-shell double boxes
\cite{HS,CGR} (see also Chapter~5 of \cite{S4}).}
It is based on the MB representation
\bea
\frac{1}{(X+Y)^{\lm}} = \frac{1}{\Gm(\lm)} \frac{1}{2\pi  i}
\int_{- i  \infty}^{+ i  \infty} \frac{Y^z}{X^{\lm+z}}
\Gm(\lm+z) \Gm(-z)\, \dd z
\label{MB}
\eea
applied to replace a sum of two terms raised to some power by
their products to some powers.

\begin{figure}[htbp]
\begin{center}
\ \psfig{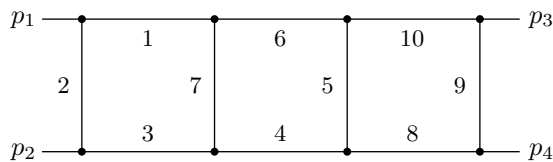}
\end{center}
\caption{Labelled triple box.}
\label{figure:tbox}
\end{figure}

\begin{figure}[htbp]
\begin{center}
\ \psfig{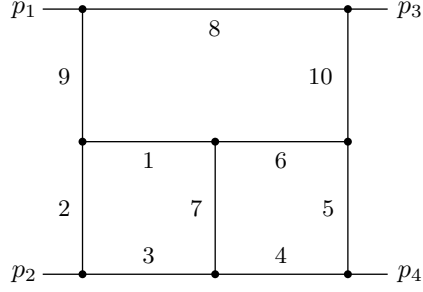}
\end{center}
\caption{Labelled tennis court.}
\label{figure:tcourt}
\end{figure}

The first step of the method is the derivation of an appropriate
MB representation. It is very desirable to do this for general
powers of the propagators (indices) and irreducible numerators.
On the one hand, this provides crucial checks of a given
MB representation using simple partial cases. (For example, one
can shrink either horizontal or vertical lines to points, i.e.
set the corresponding indices to zero, and obtain simple diagrams
quite often expressed in terms of gamma functions.)
On the other hand, such a general derivation provides unambiguous
prescriptions for choosing integration contours (see details in
\cite{S4}). So, we consider the off shell triple box and tennis
court labelled as shown in
Figs.~\ref{figure:tbox} and~\ref{figure:tcourt},
with general powers of the propagators and one irreducible
numerator in tennis court chosen as
$[(l_1+ l_3)^2]^{-a_{11}}$, where $l_{1,3}$  are the
momenta flowing through lines $1$ and $3$ in the same direction.

Experience shows that a minimal number of MB integrations for
planar diagrams is achieved if one introduces MB integrations loop
by loop, i.e. one derives a MB representation for a one-loop
subintegral, inserts it into a higher two-loop integral, etc. This
straightforward strategy provides the following 15-fold MB
representations for the dimensionally regularised off-shell triple
box and tennis court with general indices:
\bea
T_1(a_1,\ldots,a_{10};s,t,p_1^2,p_2^2,p_3^2,p_4^2;\ep)=
\frac{\left(i\pi^{d/2} \right)^3 (-1)^a(-s)^{6-a-3\ep}}{
\prod_{j=2,4,5,6,7,9}\Gm(a_j)
\Gm(4 - 2 \ep- a_{4,5,6,7} )}
&& \nn \\ && \hspace*{-125mm}
\times \frac{1}{(2\pi i)^{15}}
\int_{-i\infty}^{+i\infty}  \prod_{j=1}^{15} \dd z_j
\frac{(-p_1^2)^{z_{12}} (-p_2^2)^{z_{13}}
(-p_3^2)^{z_{4,9,14}} (-p_4^2)^{z_{5,10,15}}
(-t)^{z_{11}}}{(-s)^{z_{4,5,9,10,11,12,13,14,15}}}
\nn \\ && \hspace*{-125mm} \times
\frac{\Gm(a_9 + z_{11,12,13})
\Gm(a_7 + z_{1,2,3})
\Gm(2 - \ep- a_{5,6,7}  - z_{1,2,4})
\Gm(2 - \ep- a_{4,5,7}  - z_{1,3,5})}{ \Gm(a_1 - z_{2}) \Gm(a_3 - z_{3})
\Gm(4 - 2 \ep- a_{1,2,3}  + z_{1,2,3})}
\nn \\ &&  \hspace*{-125mm} \times
\frac{\Gm(a_5 + z_{1,4,5})
\Gm(a_{4,5,6,7} + \ep-2  + z_{1,2,3,4,5})
\Gm(z_{11,14,15} - z_{6})}{ \Gm(a_8 - z_{7}) \Gm(a_{10} - z_{8})
\Gm(4 - 2 \ep - a_{8,9,10} + z_{6,7,8})}
\nn \\ &&  \hspace*{-125mm} \times
\Gm(2- \ep - a_{8,9} + z_{6,7} - z_{11,12,14})
\Gm(2 - a_{2,3} - \ep + z_{1,3} - z_{6,8,10})
\nn \\ &&  \hspace*{-125mm} \times
\Gm(a_{8,9,10}+\ep-2 + z_{11,12,13,14,15}- z_{6,7,8})
\Gm(2- \ep - a_{9,10}  + z_{6,8} - z_{11,13,15})
\nn \\ &&  \hspace*{-125mm} \times
\Gm(a_2 + z_{6,7,8})
\Gm(2- \ep - a_{1,2}  + z_{1,2} - z_{6,7,9})
\nn \\ &&  \hspace*{-125mm} \times
\Gm(z_{6,9,10}-z_{1} )
\Gm(a_{1,2,3} + \ep-2 - z_{1,2,3}  + z_{6,7,8,9,10})
\prod_{j=2,3,4,5,7,\ldots,15} \Gm(-z_j)\;;
\label{TB}
\eea
\bea
T_2(a_1,\ldots,a_{11};s,t,p_1^2,p_2^2,p_3^2,p_4^2;\ep)=
\frac{\left(i\pi^{d/2} \right)^3 (-1)^a (-s)^{6-a-3\ep}}{
\prod_{j=2,4,5,6,7,9}\Gm(a_j)\Gm(4 - 2 \ep- a_{4,5,6,7} )}
&& \nn \\ && \hspace*{-130mm}
\times \frac{1}{(2\pi i)^{15}}
\int_{-i\infty}^{+i\infty}  \prod_{j=1}^{15} \dd z_j
\frac{ (-p_1^2)^{z_{12}} (-p_2^2)^{z_{13}}
(-p_3^2)^{z_{5,10,14}}
(-p_4^2)^{z_{15} + z_{8}}
(-t)^{z_{11}}}{(-s)^{z_{5,8,10,11,12,13,14,15}}} \prod_{j=2}^{15} \Gm(-z_j)
\nn \\ && \hspace*{-130mm} \times
\frac{\Gm(a_9 + z_{11,12,13})
\Gm(a_7 + z_{1,2,3})
\Gm(2 - a_{5,6,7} - \ep - z_{1,2,4})
\Gm(2 - a_{4,5,7} - \ep - z_{1,3,5})}{\Gm(a_1 - z_{2}) \Gm(a_3 - z_{3})
\Gm(4- 2 \ep - a_{1,2,3}  + z_{1,2,3}) \Gm(a_{10} - z_{7})}
\nn \\ && \hspace*{-130mm} \times
\frac{\Gm(a_5 + z_{1,4,5})
\Gm(a_{4,5,6,7} + \ep -2+ z_{1,2,3,4,5})
\Gm(2 - a_{2,3} - \ep + z_{1,3} - z_{6,8,10})}
{\Gm(8 - 4 \ep- a - z_{5,6,8,10})
\Gm(a_8 - z_{4,9})
\Gm(a_{1,2,3,4,5,6,7,11}+ 2 \ep-4  + z_{4,5,6,7,8,9,10})}
\nn \\ && \hspace*{-130mm} \times
\Gm(6 -a+ a_{10}- 3 \ep - z_{5,6,7,8,10,11,12,14})
\Gm(a+ 3 \ep-6 + z_{5,6,8,10,11,12,13,14,15})
\nn \\ && \hspace*{-130mm} \times
\Gm(a_2 + z_{6,7,8})
\Gm(2 - \ep- a_{1,2}  + z_{1,2} - z_{6,7,9})
\Gm(6- 3 \ep -a+a_{8}- z_{4,5,6,8,9,10,11,13,15})
\nn \\ && \hspace*{-130mm} \times
\Gm(z_{6,9,10} -z_{1})
\Gm(a_{1,2,3} + \ep-2 - z_{1,2,3} + z_{6,7,8,9,10})
\nn \\ && \hspace*{-130mm} \times
\Gm(a_{1,2,3,4,5,6,7,11}+ 2 \ep-4+ z_{4,5,6,7,8,9,10,11,14,15})\;.
\label{TC}
\eea
Here
$a_{4,5,6,7}=a_4+a_5+a_6+a_7, a=\sum a_i, z_{11,12,13}=z_{11}+z_{12}+z_{13}$,
etc. Moreover, in contrast to the rest of the paper,
the letters $s$ and $t$ denote, in these equations
as well in other equations of this section, the usual Mandelstam
variables $s=(p_1+p_2)^2$ and $t=(p_1+p_3)^2$.

These representation are written for the Feynman integrals in
Minkowski space. (This is rather convenient, in particular this allows
one to put
some of the legs on-shell.) The corresponding Euclidean versions are
obtained by the replacements $-s \to s, -t \to t, -p_1^2 \to
p_1^2,\ldots$ and by omitting the prefactors $(-1)^a$ and $i^3$.

To calculate the triple box we need, i.e.
$T_1^{(0)}=T_1(1,\ldots,1)$ at $d=4$, we simply set all the
indices $a_i$ to one. We cannot immediately set $\ep=0$ because
there is $\Gm(- 2\ep)$ in the denominator. The value of the
integral is, of course, non-zero, so that some poles in $\ep$
arise due to the integration. To resolve the structure of poles
one can apply Czakon's code \cite{Czakon}, which provides the
following value of the integral in the limit $\ep\to 0$ after
relabelling  the variables by
$z_{10} \to  z_{2}, z_{14} \to  z_{3}, z_{15} \to  z_{4}, z_{11} \to
z_{5}, z_{12} \to z_{6}$:
\bea
T_1^{(0)}=
\frac{\left(i\pi^{2} \right)^3}{(2\pi i)^{6}}
\int_{-i\infty}^{+i\infty}  \prod_{j=1}^{6} \dd z_j
\frac{ (-p_1^2)^{z_{6}} (-p_2^2)^{-1 - z_{5,6}}
(-p_3^2)^{-1 - z_{5,6}} (-p_4^2)^{z_{6}} (-t)^{z_{5}}}{(-s)^{2 - z_{5}}}
&& \nn \\ &&  \hspace*{-125mm} \times
\frac{\Gm(1 + z_{3,4}) \Gm(1 + z_{1} - z_{3,4,5})
\Gm(z_{2,3,4,5}-z_{1})\Gm(z_{4} - z_{6})}{\Gm(1 + z_{4} - z_{6}) \Gm(1 + z_{2,4} - z_{6})
\Gm(2 + z_{1,6} - z_{2,4})\Gm(2 + z_{3,5,6})}
\prod_{j} \Gm(-z_j)
\nn \\ &&  \hspace*{-125mm} \times
 \Gm(z_{2,4} - z_{6})^2
\Gm(1 + z_{1,6} - z_{2,4})^2 \Gm(1 + z_{5,6})
\Gm(1 + z_{3,5,6})\;.
\label{TB0}
\eea

To calculate the tennis court we need, i.e.
$T_2^{(0)}=T_2(1,\ldots,1,-1)$ at $d=4$, we proceed like in the
previous case. Czakon's code provides the following integral
(after relabelling $z_{10} \to  z_{2}, z_{14} \to  z_{3}, z_{15} \to  z_{4},
z_{11} \to  z_{5}, z_{12} \to  z_{6}$):
\bea
T_2^{(0)}=
\frac{\left(i\pi^{2} \right)^3}{(2\pi i)^{6}}
\int_{-i\infty}^{+i\infty}  \prod_{j=1}^{6} \dd z_j
\frac{(-p_1^2)^{z_{6}} (-p_2^2)^{-1 - z_{5} - z_{6}}
(-p_3^2)^{-1 - z_{5} - z_{6}} (-p_4^2)^{z_{6}}(-t)^{z_{5}} }{(-s)^{1 - z_{5}} }
\nn \\ &&  \hspace*{-125mm} \times
\frac{\Gm(1 + z_{3,4})
\Gm(1 + z_{1} - z_{3,4,5})  \Gm(z_{2,3,4,5}-z_{1})
\Gm(z_{4} - z_{6})}{\Gm(1 + z_{4} - z_{6}) \Gm(1 + z_{1} - z_{2,3,5,6})
\Gm(2 + z_{3,5,6}) \Gm(2 + z_{2,3,5,6})}
\prod_{j} \Gm(-z_j)
\nn \\ &&  \hspace*{-125mm} \times
\Gm(z_{1} - z_{2,3,5,6})^2
\Gm(1 + z_{5,6}) \Gm(1 + z_{3,5,6})
\Gm(1 + z_{2,3,5,6})^2 \;.
\label{TC0}
\eea

Now the simple change of variables $z_{2} \to  -z_{2} + z_{1} - z_{3}
- z_{4} - z_{5}$
in (\ref{TC0}) leads to an expression identical to (\ref{TB0}) up
to a factor of $s$ and we obtain the identity
$T_2^{(0)}= s T_1^{(0)}$, which corresponds to the identity
$\Phi^{(3)} = \Psi^{(3)}$ of the previous section. (Observe that
the factor $s$ here appears because the general integrals
(\ref{TB}) and (\ref{TC}) are defined without the appropriate
prefactors present in the definitions of $\Phi^{(3)}$ and
$\Psi^{(3)}$.

Let us stress that one can also apply the technique of MB
representation in a similar way in various situations where a given
four-point off-shell Feynman integral cannot be reduced to ladder integrals.

\subsection*{Acknowledgements}
The work of V.S. was supported by the Russian Foundation for
Basic Research through grant 05-02-17645. E.S. thanks S. Ananth,
B. Eden, Ya. Stanev, and V. Velizhanin for useful discussions and
acknowledges the warm hospitality of the Albert Einstein Institute for
Gravitational Physics in Potsdam where part of this work was done.

\end{document}